\documentstyle[sprocl]{article}

\input{psfig}
\def\be{\begin{equation}}
\def\ee{\end{equation}}
\def\bea{\begin{eqnarray}}
\def\eea{\end{eqnarray}}
\def\gsim{\mathrel{\raise.3ex\hbox{$>$}\mkern-14mu
             \lower0.6ex\hbox{$\sim$}}}
\def\lsim{\mathrel{\raise.3ex\hbox{$<$}\mkern-14mu
             \lower0.6ex\hbox{$\sim$}}}
\begin{document}

\title{BEYOND THE  STANDARD MODEL:  AN ANSWER AND TWENTY QUESTIONS}

\author{FRANK WILCZEK}

\address{School of Natural Sciences, Institute for Advanced Study\\
Princeton, NJ 08540, USA}

%\author{ A.N. OTHER }

%\address{Department of Physics, Theoretical Physics, 1 Keble Road,\\
%Oxford OX1 3NP, England}


\maketitle\abstracts{ }
  
%\section{Triumph of the Standard Model}

In dedicating this school to marking the 50th anniversary of the birth
of particle physics, Professor Zichichi posed a special challenge and
opportunity.  When we think about our subject on such a large scale,
we realize just how much was accomplished in a short time.  Fifty
years ago pions and kaons were `discovered', in the sense that
one first clearly distinguished pions from muons, and there were a
pair of cosmic ray events that seemed to indicate the existence of
unstable, heavy, hitherto unknown particles.  From today's
perspective, we can see that these discoveries were the first steps
along a path leading to completely new perspectives.  Fifty years ago,
one had the subjects of nuclear forces and beta decay.  These were
rich subjects in themselves, and for applications (nuclear reactors,
atomic weapons, theory of stars).  But the new discoveries made it
obvious that nuclear forces ultimately should not be considered a
self-contained subject, that at higher energies these `forces'
mutated into a much richer subject involving new particles and
production processes, that the subject needed a new name: the strong
interaction.  Similarly the new discoveries showed that beta decay was
just one exemplar of a wider set of instabilities infecting other
particles.  Some universality in these instabilities was perceived,
and beta decay evolved into the weak interaction.  A third great
discovery 50 years ago was the Lamb shift.  The outcome of this
discovery was quite different.  Instead of a new theory, one
discovered the unexpected, latent power of existing ideas.  By taking
quantum electrodynamics in full seriousness as a relativistic quantum
field theory, theorists were able to account for this and other
radiative corrections to the naive theory, with awesome precision.

From these first stirrings of strong and weak interaction theory, and
modern quantum field theory, to their fruition in the Standard Model was
about 25 years.  The succeeding 25 years have seen, first of all, the
consolidation of the Standard Model, after extremely extensive and
rigorous testing.  But in learning in great depth and detail what the
Standard Model can do, we also have become acutely aware of what it
can't do.  The remaining questions are especially profound, because
they have eluded an extremely powerful, successful theory.  I have
been asked to discuss physics beyond the Standard Model, with the
time scale of 50 years.  One clear lesson from the history just
discussed, is that it is foolhardy to try to do this too precisely.
But vague discussions are mostly useless.  So what I will present are
precise questions (and a few tentative answers).  I hope -- and
suspect -- that several of them will be answered, and others made to
look silly, in a significantly shorter time than 50 years.

To set the stage for the questions, let us begin with a quick
overview of the Standard Model.  The core of the Standard Model
\cite{sm,weinsalam,sutheory} of particle physics is easily displayed
in a single Figure, here Figure 1.  There are gauge groups
$SU(3)\times SU(2)\times U(1)$ for the strong, weak, and
electromagnetic interactions.  The gauge bosons associated with these
groups are minimally coupled to quarks and leptons according to the
scheme depicted in the Figure.  The non-abelian gauge bosons within
each of the $SU(3)$ and $SU(2)$ factors also couple, in a canonical
minimal form, to one another.  The $SU(2)\times U(1)$ group is
spontaneously broken to the $U(1)$ of electromagnetism.  This breaking
is parameterized in a simple and (so far) phenomenologically adequate
way by including an $SU(3)\times SU(2)\times U(1)$ $(1, 2, -{1\over
2})$ scalar `Higgs' field which condenses, that is, has a
non-vanishing expectation value in the ground state.  Condensation
occurs at weak coupling if the bare (mass)$^2$ associated with the
Higgs doublet is negative.

The fermions fall into five separate multiplets under
$SU(3)\times SU(2) \times U(1)$, as depicted in  Figure 1.
The color $SU(3)$ group acts horizontally; the weak $SU(2)$
vertically, and the hypercharges (equal to the average electric
charge) are as indicated.  Note that
left- and right-handed fermions of a single type generally transform
differently.  This reflects parity violation.  It also implies that
fermion masses, which of course connect the left- and right-handed
components, only arise upon spontaneous $SU(2)\times U(1)$ breaking.

Only one fermion family has been depicted in Figure 1; of course
in reality there are three repetitions of this scheme.
Also not represented are all the complications associated with the
masses and Cabibbo-like mixing angles among the fermions.
These masses and mixing angles are naturally accommodated
as parameters within the Standard Model, but I think it is fair to say
that they are not much related to its core ideas -- more on this below.

With all these implicit understandings and discrete choices, the core
of the Standard Model is specified by three numbers -- the universal
strengths of the strong, weak, and electromagnetic interactions.  The
electromagnetic sector, QED, has been established as an extraordinarily
accurate and fruitful theory for several decades now.  Let me now
briefly
describe the current status of the remainder of the Standard Model.

%%%%%%%%%%%%%%%%%%%%%%%%%%%%%%%%%%%%%%%%%%%%%%%%%%%%%%%%%%%%%%%%%%%
%SAMPLE FIGURE FORMAT
\begin{figure}
\centerline{\psfig{figure=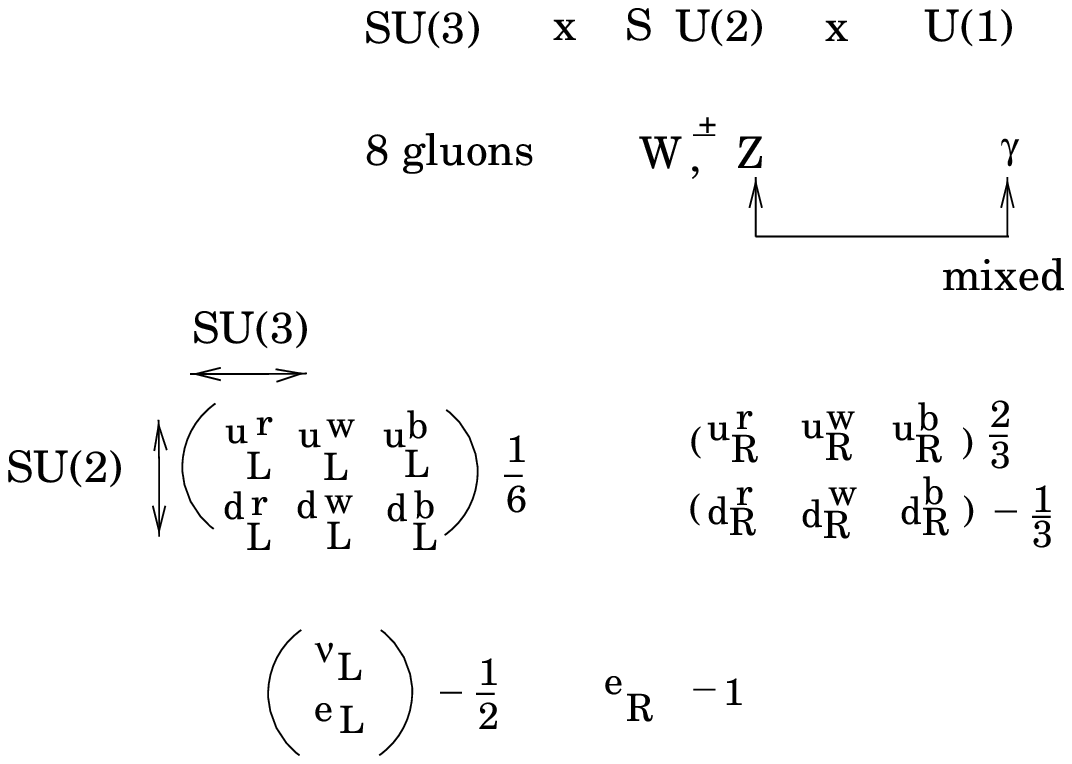,width=4in,angle=270}}
\caption[]{The core of the Standard Model: the gauge groups and the
quantum numbers of quarks and leptons.  There are three gauge groups,
and five separate fermion multiplets (one of which, $e_R$, is a singlet).
Implicit in this Figure are the universal gauge couplings -- exchanges
of vector bosons -- responsible for the classic phenomenology of the
strong, weak, and electromagnetic interactions.
The triadic replication of quark and leptons, and the Higgs field whose
couplings and condensation
are responsible for $SU(2)\times U(1)$ breaking and for fermion masses and
mixings, are not indicated.}
\label{fig1}
\end{figure}
%%%%%%%%%%%%%%%%%%%%%%%%%%%%%%%%%%%%%%%%%%%%%%%%%%%%%%%%%%%%%%%%%%%

Some recent stringent tests of the electroweak sector of the
Standard Model are summarized in Figure 2.  In general each entry represents
a very different experimental arrangement,
and is meant to test a different fundamental
aspect of the theory, as described in the caption.  There is
precisely one parameter
(the mixing angle) available within the theory, to describe all these
measurements.
As you can see, the comparisons
are generally at the level of a per cent accuracy or so.
Overall, the agreement
appears remarkably good, especially to anyone familiar with the history
of weak interactions.  

%%%%%%%%%%%%%%%%%%%%%%%%%%%%%%%%%%%%%%%%%%%%%%%%%%%%%%%%%%%%%

%\begin{figure}
%\centering
%\epsfysize=3in
%\hspace*{0in}
%\vglue-.7in
%\hglue0.75in\epsffile{LEPtable.eps}

\begin{figure}[htb]
\centerline{\psfig{figure=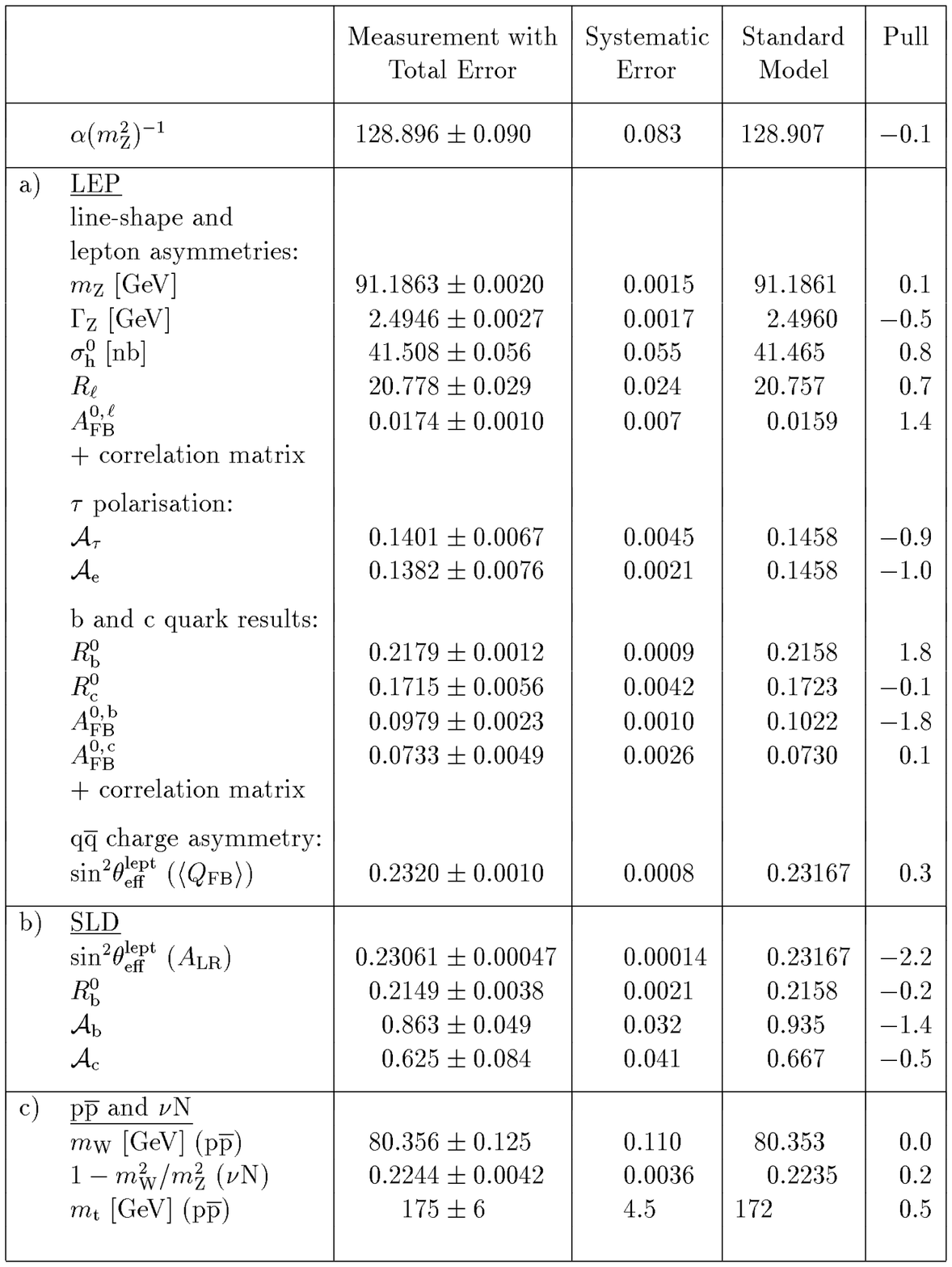,width=3.8in}}
\caption{A recent compilation of precision tests of
electroweak theory, from \protect\cite{ewfig} ,
to which you are referred for details.
Despite some `interesting' details, clearly the evidence for electroweak
$SU(2)\times U(1)$, with the simplest doublet-mediated symmetry breaking
pattern, is overwhelming.}
\label{fig2}
\end{figure}

%%%%%%%%%%%%%%%%%%%%%%%%%%%%%%%%%%%%%%%%%%%%%%%%%%%%%%%%%%%%%%

Some recent stringent tests of the strong sector of the Standard Model are
summarized in Figure 3 \cite{qcdfig}.  Again a wide variety of very
different measurements 
are represented, as indicated in the caption.  A central feature of the
theory (QCD) is that the value of the coupling, as measured in different
physical processes, will depend in a calculable way upon the characteristic
energy scale of the process.  The coupling was predicted --
and evidently is now convincingly
measured -- to decrease as the inverse logarithm
of the energy scale: asymptotic
freedom.  Again, all the experimental results must be fit with just one
parameter -- the coupling at any single scale, usually chosen as
$M_Z$.  As you can see, the agreement between theory and
experiment is remarkably good.  The accuracy of
the comparisons is at the 1-2 \% level.

%%%%%%%%%%%%%%%%%%%%%%%%%%%%%%%%%%%%%%%%%%%%%%%%%%%%%%%%%%%%%%%%%%%
%\begin{figure}
%\centering
%\vglue-.65in
%\epsfysize=3.5in
%\hspace*{0in}
%\epsffile{asdrun.eps}

\begin{figure}
\centerline{\psfig{figure=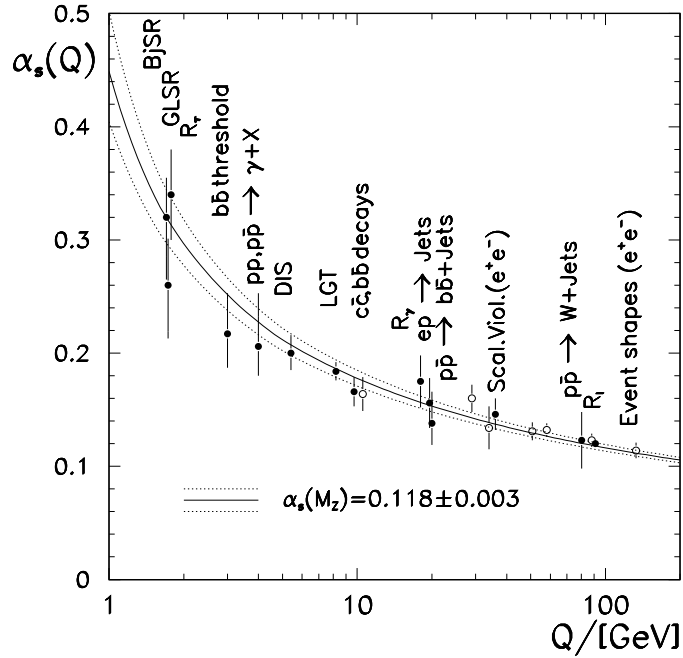,width=3.5in}}
\caption{A recent compilation of tests of QCD and asymptotic freedom, 
from \protect\cite{qcdfig},
to which you are referred for details.  Results are presented in the
form of determinations of the effective coupling $\alpha_s (Q)$ as a
function of the characteristic typical energy-momentum scale involved
in the process being measured.  Clearly the evidence for QCD in
general, and for the decrease of effective coupling with increasing
energy-momentum scale (asymptotic freedom) in particular, is
overwhelming.}
\label{fig3}
\end{figure}
%%%%%%%%%%%%%%%%%%%%%%%%%%%%%%%%%%%%%%%%%%%%%%%%%%%%%%%%%%%%%%%%%%%

Let me emphasize that these Figures barely begin to do justice to the
evidence for the Standard Model.  Several of the
results in them summarize quite
a large number of independent measurements, any
one of which might have falsified the theory.  For example, the single
point labeled `DIS' in Figure 3 describes literally hundreds of measurements
in deep inelastic scattering with different projectiles and
targets and at various energies and angles, which must
-- if the theory is correct --
all fit into a tightly constrained pattern.

The central theoretical principles of the Standard Model have been in
place for nearly twenty-five years.  Over this interval the quality of
the relevant experimental data has become incomparably better, yet no
essential modifications of these venerable principles has been
required.  Let us now praise the Standard Model:
\bigskip

\noindent$\bullet$ The Standard Model is here to stay, and describes
{\it a lot}.

Since there is quite direct evidence for each of its fundamental
ingredients ({\it i.e}. its interaction vertices), and
since the Standard Model
provides an extremely economical packaging of these ingredients, I think
it is a safe conjecture that it will be used, for the foreseeable future,
as the working description of the phenomena within its domain.  And this
domain
includes a very wide range of phenomena --
indeed not only what Dirac called
``all of chemistry and most of physics''\footnote{Dirac was referring,
here, to quantum electrodynamics.}, but also the original problems of
radioactivity and nuclear interactions which inspired the birth of particle
physics in the 1930s, and much that was unanticipated.

\bigskip

\noindent$\bullet$ The Standard Model is a {\it principled\/} theory.

Indeed, its structure embodies a few basic principles:
special relativity, locality, and quantum mechanics,
which lead one to quantum
field theories, local symmetry (and, for the electroweak sector, its
spontaneous breakdown), and renormalizability (minimal coupling).
The last of these principles, renormalizability,
may appear rather technical and perhaps less compelling
than the others; we shall shortly have occasion to re-examine it
in a larger perspective.
In any case, the fact that the Standard Model is principled in this
sense is profoundly significant: it means that its predictions are
precise and unambiguous, and generally cannot be modified `a little bit'
except in very limited, specific ways.  This feature makes the
experimental success especially meaningful, since it becomes hard to
imagine that the theory could be approximately right without in some sense
being exactly right.

\bigskip

\noindent$\bullet$ The Standard Model {\it can be extrapolated}.

Specifically because of
the asymptotic freedom property,
one can extrapolate using the Standard Model from the observed domain
of experience to much larger energies and shorter distances.  Indeed, the
theory becomes simpler -- the fundamental interactions are all
effectively
weak -- in these limits.  The whole field of very early Universe cosmology
depends on this fact, as do the impressive
semi-quantitative indications for unification and supersymmetry I shall be
emphasizing momentarily.

With this background, let me begin the questions.

\bigskip

Question 1:  Why does the Standard Model contain scattered multiplets,
with peculiar hypercharge assignments?

While little doubt can remain that the Standard Model is essentially
correct, a glance at Figure 1 is enough to show that it is not a
complete or final theory.  The fermions fall apart into five lopsided
pieces with peculiar hypercharge assignments.  This pattern needs
explanation.  Also the separate gauge theories of the strong, weak,
and electromagnetic interactions, which are conceptually and
mathematically quite similar, are practically begging to be seen as
different aspects of a more encompassing structure.

Of all the questions I will discuss, this one is outstanding because
it is the only one with a concrete and compelling answer, full of
unexpected consequences.
\bigskip

%%%%%%%%%%%%%%%%%%%%%%%%%%%%%%%%%%%%%%%%%%%%%%%%%%%%%%%%%%%%%%%%%%%%%%%
%\begin{figure}
%\parskip=0pt
%\underline{SU(5):  5 colors RWBGP}

\begin{figure}
\parskip=0pt
\parindent=40pt
\underline{SU(5):  5 colors RWBGP}

$\underline{10}$: 2 different color labels (antisymmetric tensor)

$$\matrix{\rm u_L:&\rm RP,&\rm WP,&\rm BP\cr
\rm d_L:&\rm RG,&\rm WG,&\rm BG\cr
\rm u{^c_L}:&\rm RW,&\rm WB,&\rm BR\cr
&\rm (\bar B)&\rm (\bar R)&\rm (\bar W)\cr
\rm e{^c_L}:&\rm GP&&\cr
&(\ )&&\cr
}
\pmatrix{0&\rm u^c&\rm u^c&\rm u&\rm d\cr
&0&\rm u^c&\rm u&\rm d\cr
&&0&\rm u&\rm d\cr
&*&&0&\rm e\cr
&&&&0\cr}$$

$\underline{\bar 5}$: 1 anticolor label

$$\matrix{\rm d{^c_L}:&\rm \bar R,&\rm  \bar W,&\rm  \bar B\cr
\rm e_L:&\rm \bar P&&\cr
\nu_{\rm L}:&\rm \bar G&&\cr
}
\matrix{\rm \ \ \ \ \ \ \ \ \ (d^c&\rm d^c&\rm d^c&{\rm e}&\nu)\cr}$$
\def\boxtext#1{%
\vbox{%
\hrule
\hbox{\strut \vrule{} #1 \vrule}%
\hrule
}%
%figure
}
\centerline{
\vbox{\offinterlineskip
\hbox{\boxtext{\rm Y $= -{1\over 3}$ (R+W+B) $+{1\over 2}$ (G+P)}}
}}
\caption{Organization of the fermions in one family in $SU(5)$ multiplets.
Only two multiplets are required.  In passing from this form of
displaying the gauge quantum numbers to the form familiar in the
Standard Model, one uses the bleaching rules R+W+B = 0 and G+P = 0 for
$SU(3)$ and $SU(2)$ color charges (in antisymmetric combinations).
Hypercharge quantum numbers are identified using the formula in the
box, which reflects that within the larger structure $SU(5)$ one only
has the combined bleaching rule R+W+B+G+P = 0.  The economy of this
Figure, compared to Figure 1, is evident.}
\end{figure}
%%%%%%%%%%%%%%%%%%%%%%%%%%%%%%%%%%%%%%%%%%%%%%%%%%%%%%%%%%%%%%%%%%%%%%%%

Given that the strong interactions are governed by transformations
among three color charges -- say RWB for red, white, and blue --
while the weak interactions are governed by transformations between
two others -- say GP for green and purple -- what
could be more natural than to embed both theories
into a larger theory of transformations among all five colors?
This idea has the additional attraction that an extra
U(1) symmetry commuting with the strong SU(3) and weak
SU(2) symmetries automatically appears,
which we can attempt to identify with the remaining gauge symmetry of
the Standard Model, that is
hypercharge.  For while in the separate SU(3) and SU(2) theories we
must throw out the two gauge bosons which couple respectively to
the color combinations R+W+B  and G+P, in the SU(5) theory we only
project out R+W+B+G+P, while the orthogonal
combination (R+W+B)-${3\over 2}$(G+P) remains.

Georgi and Glashow \cite{georgi74}
originated this line of thought, and showed how
it could be used to bring some order to the quark and lepton sector,
and in particular to
supply a satisfying explanation of the weird hypercharge assignments
in the Standard Model.  As shown in Figure 4, the five scattered
SU(3)$\times$SU(2)$\times$U(1) multiplets get organized into just two
representations of $SU(5)$.   It is an extremely non-trivial fact that
the known fermions fit so smoothly into $SU(5)$ multiplets.

In all the most promising unification schemes,
what we ordinarily think of as matter and anti-matter appear on
a common footing.  Since the fundamental
gauge transformations do not alter the chirality of fermions,
in order to
represent the most general transformation possibilities
one should use fields of
one chirality, say left, to represent the fermion degrees of
freedom.   To do this, for a given fermion, may require a charge
conjugation operation.
Also, of course, once we contemplate changing strong
into weak colors it will be difficult to prevent quarks and leptons
from appearing together in the same multiplets.
Generically, then,
one expects that in unified theories it will not be possible
to make a global distinction between matter and anti-matter and that
both baryon number $B$ and lepton number $L$ will be violated, as
they definitely are in $SU(5)$ and its extensions.

As shown in Figure 4, there is one group of ten left-handed fermions
that have all possible combinations of one unit of each of two
different colors, and another group of five left-handed fermions that
each carry just one negative unit of some color.  (These are the
ten-dimensional antisymmetric tensor and the complex conjugate of the
five-dimensional vector representation, commonly referred to as the
``five-bar''.)  What is important for you to take away from this
discussion is not so much the precise details of the scheme, but the
idea that {\it the structure of the Standard Model, with the particle
assignments gleaned from decades of experimental effort and
theoretical interpretation, is perfectly reproduced by a simple
abstract set of rules for manipulating symmetrical symbols}.  Thus,
for example, the object RB in this Figure has just the strong,
electromagnetic, and weak interactions we expect of the complex
conjugate of the right-handed up-quark, without our having to instruct
the theory further.  If you've never done it I heartily recommend to
you the simple exercise of working out the hypercharges of the objects
in Figure 4 and checking against what you need in the Standard Model
-- after doing it, you'll find it's impossible ever to look at the
standard model in quite the same way again.

Although it would be inappropriate to elaborate the necessary group theory
here, I'll mention that there is a beautiful extension of $SU(5)$ to
the slightly larger group $SO(10)$, which permits one to unite
all the fermions
of a family into a single multiplet \cite{georgi75}.  In fact the relevant
representation for the fermions is a 16-dimensional spinor representation.
Some of its features are depicted in Figure 5.  The 16th member of a family
in $SO(10)$, beyond the 15 familiar degrees of freedom with a Standard Model
family, has the quantum numbers of the right-handed neutrino $N_R$.

%%%%%%%%%%%%%%%%%%%%%%%%%%%%%%%%%%%%%%%%%%%%%%%%%%%%
\begin{figure}
\parskip=0pt
\hglue0.75in\underline{SO(10): 5 bit register}
\vglue-.15in
$$(\pm \pm \pm \pm \pm)\ \ :\ \ \underline{\rm even}\ \  \# \  of\  -$$
$$10:\matrix{(++-|+-)&6&\rm (u_L,d_L)\cr
(+--|++)&3&\rm u{^c_L}\cr
(+++|--)&1&\rm e{^c_L}\cr}$$

$$\bar 5:\matrix{(+--|--)&\bar 3&\rm d{^c_L}\cr
(---|+-)&\bar 2&{\rm (e_L},\nu_L)\cr}$$
\nopagebreak
$$1:\matrix{(+++|++)&1&\rm N_R\cr}$$

\caption{Organization of the fermions in one family, together with a
right-handed neutrino degree of freedom, into a single multiplet under
$SO(10)$.  The components of the irreducible spinor representation,
which is used here, can be specified in a very attractive way by using
the charges under the $SO(2)\otimes SO(2)\otimes SO(2)\otimes
SO(2)\otimes SO(2)$ subgroup as labels.  They then appear as arrays of
$\pm$ signs, resembling binary registers.  There is the rule that one
must have an even number of - signs.  Strong $SU(3)$ acts on the first
three components, weak $SU(2)$ on the final two.  The $SU(5)$ quantum
numbers are displayed in the left-hand column, the number of entries
with each sign-pattern just to the right, and finally the usual
Standard Model designations on the far right.}
\end{figure}
%%%%%%%%%%%%%%%%%%%%%%%%%%%%%%%%%%%%%%%%%%%%%%%%%%%

We have seen that simple unification schemes are successful at the
level of
{\it classification}; but new questions arise when we consider the
dynamics which underlies them.

Part of the power of gauge symmetry is that it fully dictates the
interactions of the gauge bosons, once an overall coupling constant
is specified.  Thus if SU(5) or some higher symmetry were exact, then
the fundamental
strengths of the different color-changing interactions would have
to be equal, as would the
(properly normalized) hypercharge coupling strength.  In reality the
coupling strengths of the gauge bosons in SU(3)$\times$SU(2)$\times$U(1)
are observed not to be equal, but rather to follow the pattern
$g_3 \gg g_2 > g_1$.

Fortunately, experience with QCD emphasizes that couplings ``run''.
The physical mechanism of this effect is that in quantum field theory
the vacuum must be regarded as a polarizable medium, since virtual
particle-anti-particle pairs can screen charge.  Thus one might expect
that effective charges measured at shorter distances, or equivalently
at larger energy-momentum or mass scales, could be different from what
they appear at longer distances.  If one had only screening then the
effective couplings would grow at shorter distances, as one penetrates
deeper inside the screening cloud.  However it is a famous fact 
\cite{sutheory} that due to paramagnetic
spin-spin attraction of like charge vector gluons \cite{nielsen81},
these particles tend
to {\it antiscreen\/} color charge, thus giving rise to
the opposite effect -- asymptotic freedom --
that the effective coupling tends to shrink at short distances.  This
effect is the basis of all perturbative QCD phenomenology, which is a vast
and vastly successful enterprise, as we saw
in Figure 3.

For our present purpose of understanding
the disparity of the observed couplings, it is just what the doctor
ordered.
As was first pointed out by Georgi, Quinn, and Weinberg \cite{quinn74},
if a
gauge symmetry such as SU(5) is spontaneously broken at some very short
distance then we should not expect that the effective couplings probed at
much larger distances, such as are actually measured at practical
accelerators, will be equal.  Rather they will all have been affected
to a greater or lesser extent by vacuum screening and anti-screening,
starting from a common value at the unification scale but then diverging
from one another at accessible accelerator scales.
The pattern $g_3 \gg g_2 > g_1$ is just what one should
expect, since the antiscreening or asymptotic freedom effect is more
pronounced for larger gauge groups, which have more types of virtual
gluons.

The marvelous thing is that the running of the couplings
gives us a truly quantitative
handle on the ideas of unification, for the following reason.  To fix
the relevant aspects of unification, one basically needs
only to fix two parameters: the scale at which the couplings unite, which
is essentially the scale at which the unified symmetry breaks; and
their value when they unite.  Given these, one calculates three outputs:
the three {\it a priori\/} independent couplings for the gauge groups
SU(3)$\times$SU(2)$\times$U(1) of the Standard Model.
Thus the framework is eminently
falsifiable.  The miraculous thing is, how close it comes to working
(Figure 6).

%%%%%%%%%%%%%%%%%%%%%%%%%%%%%%%%%%%%%%%%%%%%%%%%%%%%%%%%%%%%%%%%%%%
%\begin{figure}
%\centering
%\epsfysize=3.5in
%\hspace*{0in}
%\vglue-0.75in
%\hglue0.50in\epsffile{cosmo6.EPS}

\begin{figure}
\centerline{\psfig{figure=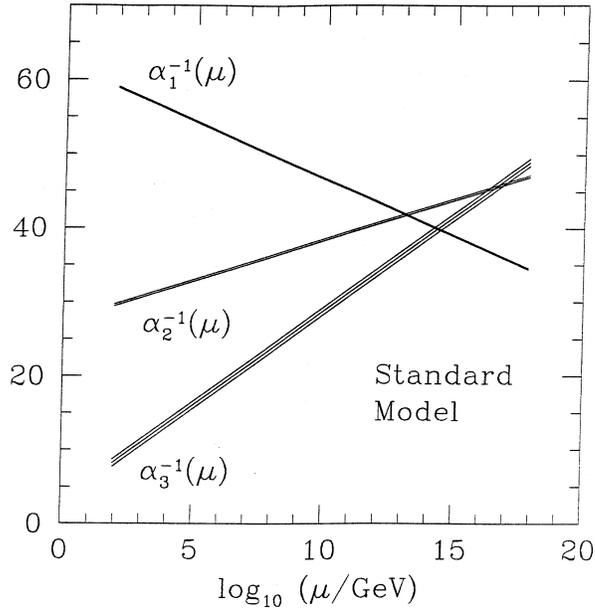,width=3.25in}}
\caption{Evolution of Standard Model effective (inverse) couplings toward small
space-time distances, or large energy-momentum scales.  Notice that
the physical behavior assumed for this Figure is the direct
continuation of Figure 3, and has the same conceptual basis.  The
error bars on the experimental values at low energies are reflected in
the thickness of the lines.  Note the logarithmic scale.  The
qualitative aspect of these results is extremely encouraging for
unification and for extrapolation of the principles of quantum field
theory, but there is a definite small discrepancy with recent
precision experiments.}
\label{fig6}
\end{figure}
%%%%%%%%%%%%%%%%%%%%%%%%%%%%%%%%%%%%%%%%%%%%%%%%%%%%%%%%%%%%%%%%%%%

The unification of couplings occurs at a very large mass scale,
$M_{\rm un.} \sim 10^{15}~{\rm Gev}$.  In the simplest version, this
is the magnitude of the scalar field vacuum expectation value that
spontaneously breaks SU(5) down to the
standard model symmetry SU(3)$\times$SU(2)$\times$U(1),
and is analogous to the scale $v \approx 250~ {\rm
Gev}$ for electroweak symmetry breaking.  The largeness of
this large scale mass scale
is important in several ways:

$\bullet~$ It
explains why the exchange of gauge bosons that are in SU(5) but not
in SU(3)$\times$SU(2)$\times$U(1), which re-shuffles
strong into weak colors
and generically violates the conservation of baryon number,
does not lead to a catastrophically quick decay of nucleons.  The rate
of decay goes as the inverse fourth power of the mass of the exchanged
gauge particle, so the baryon-number violating processes are predicted to
be far slower than ordinary weak processes, as they had better be.

$\bullet~$ $M_{\rm un.}$ is significantly smaller than the Planck scale
$M_{\rm Planck} \sim 10^{19}~{\rm Gev}$ at which exchange of gravitons
competes quantitatively with the other interactions, but not ridiculously
so.  This indicates that while the unification of couplings calculation
itself is probably safe from gravitational corrections, the unavoidable
logical next step in unification must be to bring gravity into the mix.

$\bullet~$ Finally one must ask how the tiny ratio of
symmetry-breaking mass scales $v/M_{\rm un.} \sim 10^{-13}$
required arises dynamically, and whether it is stable.  This is the
so-called gauge hierarchy problem, which I shall discuss in a
more concrete
form momentarily.

The success of the GQW calculation in
explaining the observed hierarchy $g_3 \gg g_2 > g_1$ of
couplings and the approximate stability of the proton is quite
striking.
In performing it, we assumed that the known and
confidently expected particles of the Standard Model exhaust
the spectrum up to the unification scale, and that the
rules of quantum field
theory could be extrapolated without alteration
up to this mass scale -- thirteen orders
of magnitude beyond the domain they were designed to describe.
It is a triumph for minimalism, both existential and conceptual.

%%5

However, on further examination
it is not quite good enough.  Accurate modern measurements
of the couplings show a small but definite discrepancy between the
couplings, as appears in Figure 6.  And heroic dedicated experiments to
search for proton decay did not find it \cite{blewitt85}; they currently
exclude the minimal SU(5) prediction
$\tau_p \sim 10^{31}~{\rm yrs.}$ by about two orders of magnitude.

Given the scope of the extrapolation
involved, perhaps we should not have
hoped for more.
There are several perfectly plausible bits of physics
that could upset the calculation, such as the existence of particles
with masses much higher than the electroweak but much smaller than the
unification scale.  As virtual particles these would affect the running
of the couplings, and yet one
certainly cannot exclude their existence on direct experimental
grounds.  If we just add particles in some haphazard
way things will
only get
worse: minimal SU(5) nearly works, so the generic perturbation
from it will be deleterious.  This is a major difficulty for so-called
technicolor models, which postulate many new particles in
complicated patterns.
Even if some {\it ad hoc\/}
prescription could be made to work,
that would be a disappointing outcome from what
appeared to be one of our most precious, elegantly
straightforward clues regarding physics well
beyond the Standard Model.

Fortunately, there is a theoretical idea which is attractive in many
other ways, and seems to point a way out from this impasse.  That is
the idea of supersymmetry \cite{ferrara86}.  Supersymmetry is a symmetry that 
extends
the Poincare symmetry of special relativity
(there is also a general relativistic version).  In a supersymmetric
theory one has not only
transformations among particle states with different energy-momentum but
also between particle states of different {\it spin}.  Thus spin 0
particles can be put in multiplets together with spin ${1\over 2}$
particles, or spin ${1\over 2}$ with spin 1, and so forth.

Supersymmetry is certainly not a symmetry in nature: for example, there
is certainly no bosonic particle with the mass and charge of the electron.
More generally if one defines the $R$-parity quantum number
$$
R~\equiv~ (-)^{3B+L+2S}~,
$$
which should be accurate to the extent that baryon and lepton number are
conserved, then one finds that all currently known particles are
$R$ even whereas their supersymmetric partners would be $R$ odd.
Nevertheless there are
many reasons to be interested in supersymmetry, and especially in
the hypothesis that supersymmetry is effectively broken at a relatively
low scale, say $\approx$ 1 Tev.  Anticipating this for the moment, let
us consider the consequences for running of the couplings.

The effect of low-energy supersymmetry on the running of the couplings
was first considered long ago \cite{dimopoulos81},
well before the discrepancy described above
was evident experimentally.
One might
have feared that such a huge expansion of the theory, which essentially
doubles the spectrum, would utterly destroy the approximate success of
the minimal SU(5) calculation.  This is not true, however.  To a first
approximation, roughly speaking because it 
is a space-time as opposed to an internal
symmetry,  supersymmetry does not affect the group-theoretic structure of the
unification of couplings calculation.  
The absolute
rate at which the couplings run
with momentum is affected, but not the relative rates.  The main effect
is that the supersymmetric partners of the color gluons, the gluinos,
weaken the asymptotic freedom of the strong interaction.  Thus they
tend to
make its effective
coupling decrease and approach the others more slowly.  Thus
their merger requires a longer lever arm,
and the scale at which the couplings meet increases by an order of
magnitude or so, to about 10$^{16}$ Gev.
Also the common value of the effective couplings at
unification is slightly larger than in conventional unification
(${g^2_{\rm un.} \over 4\pi } \approx {1\over 25}$ {\it versus\/}
${1\over 40}$).  This increase in unification scale
significantly reduces the predicted rate for proton decay
through exchange of the dangerous color-changing gauge bosons,
so that it no longer conflicts with existing
experimental limits.

Upon more careful examination there is another effect of low-energy
supersymmetry on the running of the couplings, which although quantitatively
small has become of prime interest.  There is an important exception to
the general rule that adding supersymmetric partners does not immediately
(at the one loop level)
affect the relative rates at which the couplings run.  This
rule works for particles that come in complete SU(5) multiplets, such as
the quarks and leptons (which, since they don't
upset the full SU(5) symmetry, have basically no effect)
or for the supersymmetric partners of the
gauge bosons, because they just renormalize the existing,
dominant effect of the
gauge bosons themselves.  However there is one peculiar additional
contribution, from the supersymmetric partner of the Higgs doublet.
It affects only the weak SU(2) and hypercharge U(1) couplings.
(On phenomenological grounds the
SU(5) color triplet partner of the Higgs doublet must be extremely massive,
so its virtual exchange is not important below the unification scale.
{\it Why\/}
that should be so, is another aspect of the hierarchy problem.)
Moreover, for slightly technical reasons even in the
minimal supersymmetric model it is necessary to have
two different Higgs doublets with opposite hypercharges\footnote{Perhaps
the simplest, though not the
most profound, way to appreciate the reason for this has to do with
anomaly cancelation.  The minimal
spin-1/2 supersymmetric partner of the
Higgs doublet is chiral and has non-vanishing hypercharge, introducing
an anomaly.  By including a partner for the anti-doublet, one cancels
this anomaly.}.
The main effect of
doubling the number of Higgs fields and including their supersymmetric
partners is a sixfold enhancement of the asymmetric
Higgs field contribution to
the running of weak and hypercharge couplings.  This causes a
small, accurately calculable change in the calculation.
{}From Figure 7 you see that it is a most welcome one.   Indeed,
in the minimal
implementation of supersymmetric unification, it puts the running of
couplings calculation right back on the money \cite{ellis91}.

%%%%%%%%%%%%%%%%%%%%%%%%%%%%%%%%%%%%%%%%%%%%%%%%%%%%%%%%%%%%%%%%%%%
%\begin{figure}
%\centering
%\epsfysize=3.5in
%\hspace*{0in}
%\vglue-.75in
%\hglue0.65in\epsffile{cosfig7.ps}

\begin{figure}
\centerline{\psfig{figure=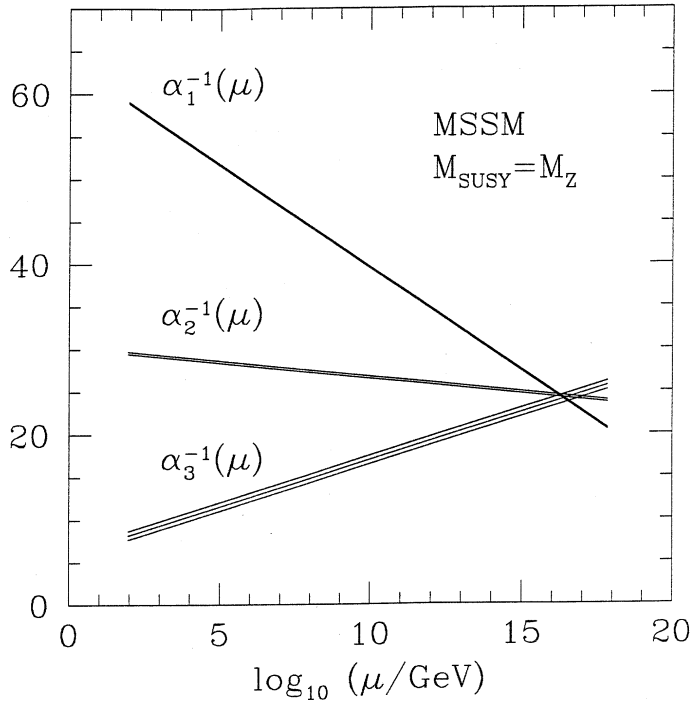,width=3.5in}}
\caption{Evolution of the effective (inverse) couplings in the minimal
extension of the Standard Model, to include supersymmetry.  The
concepts and qualitative behaviors are only slightly modified from
Figure 6 (a highly non-trivial fact!) but the quantitative result is
changed, and comes into adequate agreement with experiment.  I would
like to emphasize that results along these lines were published well
before the difference between Figure 6 and Figure 7 could be resolved
experimentally, and in that sense one has already derived a successful
{\it prediction\/} from supersymmetry.}
\label{fig7}
\end{figure}
%%%%%%%%%%%%%%%%%%%%%%%%%%%%%%%%%%%%%%%%%%%%%%%%%%%%%%%%%%%%%%%%%%%

Since the running of the couplings with scales depends only
logarithmically on the mass scale, the
unification of couplings calculation is not
terribly
sensitive to the precise scale at which supersymmetry is broken,
say between 100 Gev and 10 Tev.  (To avoid confusion later, note that
here by 
``the scale at which
supersymmetry is broken''  I mean the typical mass splitting between 
Standard Model particles and their supersymmetric partners.
The phrase is frequently used in a different sense, referring to
the largest splitting between supersymmetric partners in the entire
world-spectrum; this could be much larger, and indeed in popular models
it almost invariably is.  The ambiguous terminology is endemic in the
literature; fortunately, the meaning is usually clear from the context.)
There have
been attempts to push the calculation further, in order
to address this question of
the supersymmetry breaking scale, but they are controversial.
For example, comparable uncertainties arise from the
splittings among the
very large number of particles with masses of order the unification scale,
whose theory is poorly developed and unreliable.

%[B]

Let me summarize, now, this long answer to our first question.

%xxxbullets 

$\bullet$ The unification of strong and weak color charges works
beautifully, at the level of symmetry.

$\bullet$ The ugly ducklings of the Standard Model, the hypercharges of the
fermions, become elegant swans in unified theories.

$\bullet$ The superficially most severe problems facing this
unification: the inequality of observed coupling strengths, and the
existence of baryon-number violating processes with substantial
strength, are removed, or at least drastically tempered, by careful
attention to the running of couplings.  

$\bullet$ This points to a very high scale where the unified symmetry is
spontaneously broken.

$\bullet$ The unification of couplings does not quite work out for the minimal
Standard Model.

$\bullet$ On the other hand, if we augment the minimal Standard Model
to include low-energy supersymmetry, there is stunning agreement with
experiment.

$\bullet$ Low-energy supersymmetry is attractive in several other ways;
especially, it can address the hierarchy problem. (see Figure 8)

All this provides, in my opinion, very good reasons to be optimistic
about the future of experimental particle physics at the high energy
frontier.  For much more information on supersymmetry, I heartily
recommend a lively recent review by Dienes and Kolda \cite{dienes}

%%%%%%%%%%%%%%%%%%%%%%%%%%%%%%%%%%%%%%%%%%%%%%%%%%%%%%%%%%%%%%%%%%%
%\begin{figure}
%\centering
%\vglue-.75in
%\epsfysize=3.5in
%\hspace*{0in}
%\epsffile{cosfig8.ps}

\begin{figure}
\centerline{\psfig{figure=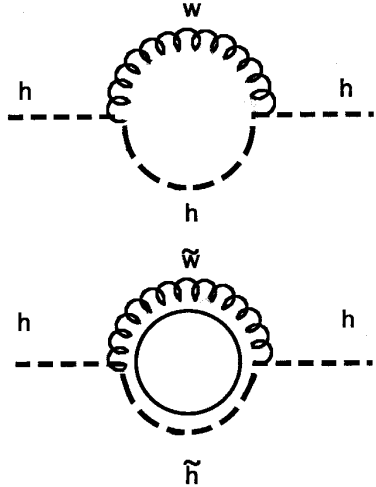,width=2in}}
\caption{Contributions to the Higgs field self-energy.  These graphs give
contributions to the Higgs field self-energy which separately are
formally quadratically divergent, but when both are included the
divergence is removed.  In models with broken supersymmetry a finite
residual piece remains.  If one is to obtain an adequately small
finite contribution to the self-energy, the mass difference between
Standard Model particles and their superpartners cannot be too great.
This -- and essentially only this -- motivates the inclusion of
virtual superpartner contributions in Figure 7 beginning at relatively
low scales.}
\label{fig8}
\end{figure}
%%%%%%%%%%%%%%%%%%%%%%%%%%%%%%%%%%%%%%%%%%%%%%%%%%%%%%%%%%%%%%%%%%%

\bigskip
Question 2:  Can gravity be brought into the unification?

A remarkable consequence of the unification of couplings calculation
is the emergence of a very large mass scale, comparable to the Planck
scale 
\begin{equation} M_{\rm Planck} ~=~ \sqrt{\hbar c \over G}
\end{equation} 
which is the value of the cutoff, such that loop
integrals involving virtual graviton exchange become of order unity.
If we had plotted the effective coupling of gravity on the same graph
as the other couplings, it would start much smaller (or with much
larger inverse coupling), evolve as a power rather than
logarithmically, and -- since the unified coupling is slightly smaller
than unity -- meet the others slightly below the Planck scale.

I said the scales were `comparable', but actually the unification
scale is 2-3 orders of magnitude smaller.  This difference is
important for the self-consistency of the calculation, since
gravitational effects were ignored.  But it is close enough to force
the question, always of course present in principle, of getting a
fully unified theory including all four interactions.  One of the
first challenges such a theory should meet, is to account for this
residual scale discrepancy.

String theory, or its elaboration into M theory, currently supplies
the best ideas for how a fully unified theory might be achieved.  This
brings up the next question ...

\bigskip
Question 3: Can string/M theory be made algorithmic? testable? user-friendly?

Professor Green told you about the latest developments in this
impressive and dynamic subject.  One cannot help but be dazzled by the
wealth of new concepts and results.  There is so much structure and
mathematical consistency that one comes to feel, as Einstein said when
asked whether he thought general relativity might be wrong (before the
measurement of the bending of light) ``then the Lord will have
missed a great opportunity.''

Yet very fundamental questions and limitations remain.  They are not
just technicalities, and should not be underestimated.  My question
attempts to identify them in a precise and I hope constructive way.

A theory is {\it algorithmic\/} if one knows, in principle, how to
compute its consequences.  It may be easy to compute some
consequences, and very difficult in practice to compute others.  QCD
is an excellent example of this.  Calculating short distance
properties is easy (or at least straightforward) while calculating
nuclear binding energies is hard.  It may be that one can only compute
up to a certain accuracy, and the theory will break down if pushed too
hard.  QED provides an excellent example of this.  Its perturbation
expansion has given us far and away the most accurately tested results
in natural science, yet this expansion does not converge and the
theory, as normally defined and used, does not properly exist.  (More
precisely, one cannot satisfy the full axioms of relativistic quantum
field theory in using just electrons coupled to photons.  One can work
with a version of QED cut off at an astronomically large energy scale,
which ultimately fails to satisfy the axioms, but has the same
practical consequences as naive QED.)  In any case, both for QCD and
for QED there is a mechanical procedure to answer all physically
meaningful questions within their scope.  On the other hand, I don't
think anyone would know how to program a computer, even in principle,
to compute whether string theory gives us, for example, exactly four
macroscopic space-time dimensions.

Another criterion is whether the theory is {\it testable}.  This is
not as clear-cut a question as it might sound.  One can imagine
different reactions to a given body of evidence -- juries are not
always unanimous.  Feynman was expressing doubts about QCD well into
the 1980s.  Certainly the discovery of supersymmetry would be
extremely encouraging for string theory, since the theory incorporates
supersymmetry in its sinews.  But on the other hand supersymmetry is
an idea that can stand on its own, and the idea that it should be
broken only at low energies is by no means an obvious consequence of
string theory (compare, in this regard, ten-dimensional Lorentz
invariance).  In any case, I think we can all agree that more
characteristic, less generic sorts of predictions would be welcome.

This brings us to the final desideratum, that the theory should become
{\it user friendly}.  Concretely, what I mean is this: we may have the
germs of an overarching theory that is supposed to contain answers to
each one of the questions I am posing here (except this one, and
number 20).  Yet I think it is fair to say that no such answers,
derived from string theory, are currently in place.  We shouldn't
have to guess whether this is due to temporary mathematical
difficulties, or something more fundamental.  We must aspire to more
meaningful contact between high theory and the major features of
observable reality beyond the Standard Model.

\bigskip
Question 4:  Why is the cosmological term so small?

We have become accustomed to the idea that the vacuum, which evolution
has encouraged us to regard as empty and void, is in reality full of
various symmetry-breaking condensates and fluctuating fields.  Yet
gravity seems to be oblivious to this structure: the energy-momentum
tensor of empty space, proportional to the cosmological term, is
either zero or at least smaller by many orders of magnitude than naive
expectations (from condensation energies, for example.)

In operator language, the cosmological term corresponds to the
interaction with the unit operator, $ \sqrt g {\cal L} = \Lambda \sqrt
g$.  No ordinary symmetry can banish such a term.  Its absence is a
profound problem.  To explain it might require symmetry principles of
a new type, or a re-examination of fundamental aspects of the
formulation of general relativity.

For a recent review, see \cite{swein89}.

\bigskip
Question 5: How is the hierarchy of electroweak and unified symmetry
breaking established?

If we believe in the picture suggested by the unification of
couplings, there is a vast disparity between two fundamental scales of
symmetry breaking.  In a broad sense this is explained by the fact
that the running of couplings -- both gauge couplings, and the Yukawa
couplings appearing in effective potentials -- is logarithmic, so the
range of energy scales over which significant changes occur can easily
be very large.  I have also mentioned how, at a very heuristic level,
supersymmetry can protect the low-energy effective interactions
(specifically, the Higgs mass) from being infected and pulled up to
the high scale.  There are a lot of complications, however, in making
this work in detail.

A very specific problem in this circle of ideas is to understand why
the Higgs doublet -- unlike any other component of the Standard Model
or its supersymmetric extension -- does not unify at low energies.  By
rights it should be part of a vector representation of $SU(5)$, with a
triplet partner.  But the triplet partner mediates proton decay, and
so on phenomenological grounds it must be extremely heavy, of order
the unification scale.  This split multiplet, you will recall, also
played a major role in making the unification of couplings come out
right in the MSSM.  The phenomenon of doublet-triplet splitting is
striking and qualitative, so it might be one of our best clues for
reconstructing the symmetry-breaking dynamics at the unification
scale.

\bigskip

Question 6: How is supersymmetry broken?

The unification of couplings calculation gives us some confidence that
supersymmetry is broken at the weak scale.  However, both the strength
and the weakness of this calculation is its relative insensitivity to
details.  Since the running of couplings is logarithmic, and the
couplings are small, only gross changes in the overall scale could
significantly alter the results.  Thus the question whether the
overall scale of breaking is 100 Gev or 10 Tev, and the question of
relative values of different squark, slepton, higgsino and gaugino
masses are left wide open, from a phenomenological point of view.

There are several different scenarios for supersymmetry breaking, with
quite different phenomenological consequences.  For an introduction, I
refer you again to \cite{dienes}.

\bigskip

Question 7: Why are there three families?  What explains the pattern of
masses and mixing?

Nobody knows.

One important possibility, that could potentially anchor the
discussion of this question, is that the top quark mass corresponds to
an infrared fixed point of the renormalization group \cite{topquark}.
That is, it could well be that any sufficiently large value of the top
quark- Higgs doublet Yukawa coupling at the unification scale flows
down to something close to the observed value at low energy.  This
could also be true of the bottom quark, if (as in supersymmetric
models) two separate Higgs fields, with significantly different values
of their expectation values, were responsible for top and bottom
masses.  The ratio of vacuum expectation values, in supersymmetric
models, is conventionally written $\tan \beta$; it will be very
interesting, in this connection, to see if $\tan \beta \gg 1$.

\bigskip

Question 8:  What are the neutrino masses and mixings?

Professor Totsuka has addressed experimental and phenomenological
aspects of this question at length, so I will confine myself to a few
theoretical remarks.

At the level of the Standard Model, neutrino masses are generated
through the unique $SU(3)\times SU(2) \times U(1)$ invariant dimension
5 interaction 
\begin{equation} {\cal L} ~=~ h^{ij}\epsilon_{\alpha
\beta} l^\alpha_{i\rho} l^\beta_{j\sigma} \phi^\rho \phi^\sigma
\end{equation} 
where i and j are family indices, $\alpha$ and $\beta$
are spinor indices, and $\rho$ and $\sigma$ are weak $SU(2)$ indices.
$l$ is the fermion doublet, $\phi$ is the Higgs doublet, and $h^{ij}$
is a coupling matrix.  Due to weak $SU(2)$ the neutrino mass term
comes together with lepton-number violating term involving charged
leptons, which however seem to be much less accessible experimentally.

It is very significant, from a conceptual point of view, that one must
go to such an elaborate interaction to generate lepton number
violation.  All $SU(3)\times SU(2)\times U(1)$ invariant terms of
dimension 4 or less involving {\it minimal\/} Standard Model fields
automatically conserve lepton number.  Because of the dimensions of
the fields, the $h^{ij}$ must have the dimensions of inverse mass.
Thus the smallness of lepton number violation can be cleanly traced to
the existence of a large mass scale, where interactions beyond the
Standard Model set in.

In supersymmetric extensions of the Standard Model, on the contrary,
there are lower dimension lepton-number violating terms.  One must
suppress them in some other way, presumably by imposing appropriate
symmetries.

At a slightly more microscopic level, one attractive mechanism for
generating neutrino masses is the `see-saw' mechanism \cite{gell-mann79}.  It
postulates the existence of one or more $SU(3)\times SU(2)\times U(1)$
singlet, space-time spinor fields $N_{i\alpha}$.  Three such fields --
one for each family -- arise naturally in $SO(10)$ unification
schemes, as we saw before.  Now one has the possible, $SU(3)\times
SU(2)\times U(1)$ invariant interactions 

\begin{equation} {\cal L}_M
~=~ M^{ij} \epsilon^{\alpha \beta} N_{i \alpha} N_{j\beta}
\end{equation} and \begin{equation} {\cal L}_\phi ~=~ k^{ij}
l^{\alpha}_{i \rho} N_{j\alpha} \phi^\rho .  
\end{equation} 

In the
first of these equations the coupling matrix $M$ has dimensions of
mass, and the interaction can be considered as a Majorana mass term
for the right-handed neutrinos $N$.  In the second of these equations
the coupling matrix $k$ is dimensionless.  This interaction has the
same structure as the interaction responsible for the masses of
ordinary quarks and leptons in the Standard Model, arising as $\phi$
acquires a vacuum expectation value.

Now if the eigenvalues of $M$ are very large compared to the weak
scale, then in analyzing processes at the weak scale we can ignore the
kinetic energy of the $N$ fields and integrate them out (solve their
field equations algebraically).  This procedure gives us a dimension 5
operator involving purely Standard Model fields, precisely of the type
we discussed earlier.  Its magnitude is, roughly speaking, inversely
proportional to the mass scale associated with $M$.

A major theoretical challenge is to get some concrete insight into the
matrices $M$ and $k$ (and thereby $h$), so that we can take proper
advantage of the brilliant, heroic work our experimental colleagues
are doing to determine the properties of neutrinos.

Although it would not be appropriate to go into greater detail here, I
would like to emphasize that there are likely to be important
connections among all the elements of the complex of problems involving quark
masses and mixing angles, CP violation, lepton and baryon number
violation, neutrino masses and mixing angles (and the supersymmetric
generalization of all these).  There are great opportunities here;
even partial insights could help correlate and guide experimental
investigations.

\bigskip

Question 9: How badly is baryon number broken?  What protects it?

A similar analysis of possible baryon-number violation reveals many
possibilities for dimension 5 and 6 (or even dimension 3 or 4)
interactions involving squarks \cite{dienes}.  One has to work very
hard to suppress them sufficiently, typically by invoking discrete
symmetries {\it ad hoc}.  Could these have a profound origin?

Conversely, the experimental search for nucleon decay should remain a
top priority.

\bigskip

Question 10: What is the origin of the observed CP violation?

Thus far, despite more than thirty years of intense searching after
the Cronin-Fitch experiment \cite{jjvr}, CP violation has still only
been observed in the K meson system.

Kobayashi and Maskawa \cite{kobay} had the brilliant idea that CP
violation would be induced, generically, by mixings among three
generations of quarks.  With three generations, but not with two, the
generalized Cabibbo-like angles characterizing the weak currents
contain a phase that cannot be removed by field redefinitions.  In
this way, they anticipated the discovery of the third generation.
However, we still do not know if they were right about the original
problem.  Their possible phase has not been measured, and we do not
know if it accounts quantitatively for CP violation in the K meson
system.

There are ambitious experimental programs being mounted to study CP
violation in B meson decays, which should decide this question.  For a
review, see \cite{nirquinn}.

\bigskip
Question 11:  What are the P, T  violating electric dipole moments?

Another sort of CP violating effect is an electric dipole moment of an
elementary particle.  An electric dipole moment for an elementary
fermion comes from the dimension-5 interaction 
\begin{equation} 
{\cal L}_{\rm edm} ~=~ \kappa \bar f \gamma_5 \sigma^{\mu \nu} f F_{\mu\nu}.
\end{equation} 
It is odd under both P and T.  Supersymmetric
extensions of the Standard Model contain many additional potential
sources of P and T violation, besides the phase in the weak current
that (as we mentioned) is plausibly, though not surely, responsible
for the CP violation in the K meson system.

An excellent review of both the theoretical situation and the
experimental prospects is given in Barr \cite{barr93}.

It is noteworthy that some of the most sensitive tests involve
searching for electric dipole moments of atoms and molecules, as
opposed to what we ordinarily think of as elementary particles.  Since
this is a school, I'm going to take half a minute to clear up a
point that several people have found confusing, and asked me about.
``Wait a minute'', they say, ``didn't I read in my chemistry
textbook that water molecules have an electric dipole moment? Surely
this doesn't signal fundamental symmetry breaking!''  And of
course it doesn't.  The point is that the true dipole moment, in
our sense, is defined by a strictly energy shift in response to an
electric field, $\Delta \varepsilon \propto E\cdot \langle J \rangle
$, where the angular momentum $J$ is the only available vector.  Since
the matrix element is manifestly unnatural under $P$, and $T$ odd, a
nonvanishing shift of this type does indeed violate these symmetries.
For a water molecule, however, there are very low-lying rotational
states, with opposite $P$ and $T$, that will mix with the nominal
ground state even for quite weak fields.  A non-vanishing matrix
element of $J$ between such states need violate no symmetries.  It is
a simple exercise in near-degenerate perturbation theory to show that
under these conditions the energy shift behaves as $\Delta \varepsilon
\propto \sqrt{E^2 + \delta^2}$, with a very small $\delta$.  Evidently
this can mimic, for practical purposes if the electric fields are not
too weak, the response of a true, fundamental dipole.

\bigskip
Question 12:  How is the strong P, T problem solved?  Do axions exist?  
 
I lectured on these subjects at length at a previous Erice school
\cite{fw85}.  The main development since then is that experiments
capable of seeing axionic dark matter, for an interesting range of
couplings, are underway \cite{cosmic}.  It remains true, I think, that
there is no other comparably attractive solution to the problems that
Peccei-Quinn symmetry, and the concomitant axions, address.

\bigskip

Quesion 13: Why is the $\mu$ term what it is?

This is a rather more specialized technical problem than most on my
list, but quite fundamental.  In supersymmetric models of particle
physics one needs not one but two Higgs doublets.  This is because 
supersymmetry assigns a definite character to fields, holomorphic or
antiholomorphic, and only holomorphic fields occur in the
superpotential.  Thus the trick of using the complex conjugate of the
doublet to supply an order parameter with the opposite hypercharge,
used in the minimal Standard Model, will not work in a supersymmetric
theory.  One needs two separate doublets to give masses to up and down
quarks.  A more down-to-earth explanation, is that the doublets have
fermionic partners, whose potential anomalies need to be cancelled.

In any case, a cross-coupling between the two doublets in the
superpotential is allowed by supersymmetry.  It has the dimensions of
mass.  There would seem to be two natural options: either there is a
symmetry which forbids the coupling, or it acquires (like the
right-handed neutrinos we discussed before) an extremely large mass.
The first option leads to the original weak-scale axion, which is a
very pretty possibility, but excluded by experiment.  The second
option effectively obliterates the weak scale: the Higgs doublets are
removed to the high mass scale, or (worse) electroweak symmetry is
broken at the high scale.  So it is a serious problem, why this
so-called $\mu$ term is of order the electroweak scale.  It endangers
the whole idea of weak-scale supersymmetry.

There are some promising ideas to address this problem.  They exploit
details of supergravity.  So the $\mu$ problem may turn out to provide
a particularly clean window into the unification of gravity with the
other interactions.  For more information see \cite{dienes}.

\bigskip

Question 14:  Are there additional macroscopic forces?

A variety of ideas in particle physics suggest the existence of very
light, very weakly coupled spin-0 particles.  We have already
discussed axions.  Other possibilities appearing in the literature
include exact or approximate Nambu-Goldstone fields associated with
family symmetries (familons) or, closely related to these, so-called
moduli fields, and dilatons.

Exchange of such particles, if they exist, will induce forces whose
range is the Compton wavelength of the particle.  Thus $10^{-5}$ eV
particles will give forces which die off as a power at distances
shorter than about 1 cm, then exponentially.  This sort of mass scale,
or slightly smaller, arises from the combination 
\begin{equation} \mu
\sim M_W^2 / M_{\rm Planck} 
\end{equation} 
that occurs in many
speculations.  Of course, true Nambu-Goldstone bosons would mediate
forces of infinite range.  They give a $1/r^3$ potential.

I discussed these matters in considerable detail at a previous Erice
school, \cite{fw85}.  See also the excellent experimentally oriented
review by Adelberger, et al \cite{adel91}.
Much more could be done with this subject, especially on the
experimental side.

\bigskip

Question 15:  How (and how well) are flavor symmetries protected,
outside the Standard Model?

Neutrino masses, probed in oscillation experiments, provide one window
into lepton-number violating processes, as we have already discussed.
Rare decays such as $\mu \rightarrow e \gamma$ provide another.
Supersymmetric theories introduce many possibilities for additional
lepton-number violating interactions.

Indeed, there is a big problem in understanding why mixings among the
various sparticles do not induce unacceptably large lepton-number
violation and flavor-changing processes among quarks \cite{dienes}. 
(Dienes-Kolda)

A long-shot, but very informative if true, is the possibility that
very light particles associated with spontaneous breaking of flavor
symmetries (familons) could be produced in rare decays such as $K
\rightarrow \pi f$.
 
\bigskip
Question 16:  Can we provide foundations for, or alternatives to, inflation?

Professor Linde gave us inspired lectures on inflationary models.  So
you know how appealing the basic ideas are.  However, we should
remember that the foundation of facts which support these speculations,
although profound, is extremely limited.  The main results -- spatial
flatness (or maybe not?), scale invariant fluctuation spectrum (maybe
not?)  -- are simple and qualitative, so that it does not seem
ridiculous to imagine that some completely different idea could
explain them.  Of course, one must produce the idea!  Or, one must
give names and faces to the postulated inflation fields, explain the
required flat potentials, and properly ground inflation in our
knowledge of the laws of physics.

\bigskip
Question 17:  What is the `dark matter' of cosmology?

The evidence for a dominant non-baryonic component to the mass of the
Universe has become overwhelming.  Particle physics provides two
excellent candidates: the lightest supersymmetric (R-parity odd)
particle or LSP, and axions.  Neutrinos could also provide a
significant fraction, if their mass is of order a few eV.  A
cosmological term could be considered another, most unusual, form of
dark matter.  Future measurements of the cosmic microwave anisotropy
will make it clear what astronomers need; and there are promising
experimental programs to check out the individual candidates.  For a
review, see \cite{gjung}.

\bigskip

Question 18: Why is there a cosmic asymmetry between matter and antimatter?

The general idea, that baryon-number violating processes operating in
Big Bang conditions could generate the asymmetry between matter and
antimatter, starting from symmetric conditions, has been well
established.  Many variations on the theme have appeared in the
literature.  Early work especially focussed on processes near the
unification scale; more recently there has been much work on the
possibility of generating the asymmetry at the weak scale.  Since
there are so many possibilities, most of which are rather well
insulated from laboratory experimentation, this important problem
might only be elucidated as a by-product of better understanding of
particle physics over a broad front.  For a review, see \cite{cohen}.

\bigskip

Question 19: Are the laws of physics historically determined?  Are
they the same everywhere?

Spontaneous symmetry breaking is a prominent feature of the Standard
Model.  In models of unification, there are typically several layers
of spontaneous symmetry breaking.  One can easily construct models in
which there are several local minima, with different values of the
condensed fields and thus several different sets of `laws of
physics'.  This is especially common in supersymmetric models.
Indeed, in that context it is common to have continuous so-called
moduli fields, which parametrize different zero-energy, vacuum states.
String theory, it appears, also supports many physically inequivalent
solutions.

Even if the ground state were to be in some sense unique, in any of
these contexts, one would face the question whether we could live in a
metastable vacuum, either a local minimum or with slowly varying
`constants of nature'.  These could be realized, concretely, as
very weakly coupled fields.  We are already familiar with one example
of this kind, of course -- the gravitational field.

These considerations suggest that the problem of vacuum selection --
which, in the sense used here, is an integral part of deriving the
laws of nature -- may not have a unique, universal solution, even
supposing that there is a unique, universal fundamental theory.  It
could well be that the constants of nature we experience are
determined accidentally, and differ from those in other parts of the
Universe.  This idea is very much encouraged by inflationary models,
since they readily explain how the Universe could be homogeneous on
very large scales, though ultimately extremely inhomogeneous.

\bigskip

Question 20:  Are there essentially new {\it phenomena\/} within the
Standard Model?

This is, as it stands, an open-ended question whose answer is obviously
`yes', since the Standard Model implicitly (very implicitly!)
contains chemistry and condensed matter physics as subtheories.  The
spirit of it, of course, is to ask whether there are essentially new
phenomena involving core concepts of the Standard Model in a
reasonably direct way.  I think the answer is still very much yes.

Professor Satz has told you about recent exciting developments around
the quark-gluon plasma.  The whole subject of phase transitions in QCD
is opening up from several angles: theory, numerical experiments, and
now physical experiments.  There is good reason to think that
confinement is abolished, and chiral symmetry restored, at high
temperature.  There is good reason to think that both color and
strangeness are spontaneously broken at densities a few times nuclear.
Figuring out how these changes occur is potentially important for
understanding neutrons stars and experiments using heavy ion
collisions; and in any case it is an integral part of understanding QCD
properly.

As we learn more about the world at and above the weak scale, it will
be fascinating to understand the transition or transitions that
occurred as electroweak symmetry was broken in the early Universe.
Both inflation and genesis of matter-antimatter asymmetry have been
ascribed to events during this epoch, in slight extensions of the
Standard Model.

The central mission of particle physics has always been to find the
fundamental laws of nature.  This is surely a grand goal.  But we
should also recognize and take pleasure in the intrinsic beauty of a
phenomenon like color coherence \cite{nature}, where fundamental
concepts -- in this case, the quantum interference of color fields --
are cleverly isolated and demonstrated in concrete experimental
realities.

\bigskip

Clearly, fifty years into particle physics there is no shortage of
important, interesting open questions.  Your homework assignment is to
answer one -- or, for extra credit, more than one.

\bigskip

%%%%%%%%%%%%%%%%%%%%%%%%%%%%%%%%%%%%%%%%%%%%%%%%%%%%%%%%%%%%%%%%%%%%%%%%%%
\section*{Acknowledgments}
I thank Keith Dienes for supplying
Figures 6 and 7.\\ This paper is from lectures given at the Ettore Majorana
Summer School, Erice, Italy, August 1997.  F.W. is supported in part
by DOE grant DE-FG02-90ER40542


\noindent

\begin{thebibliography}{99}
%1
\bibitem{sm}
For recent reviews of the Standard Model, see References 4 and 5.

%2
\bibitem{weinsalam}
After several partial and tentative proposals, the $SU(2)\times U(1)$
electroweak theory took on in its essentially modern form in:
S. Weinberg, Phys. Rev. Lett. {\bf 19}, 1264 (1967); 
A. Salam, in {\it Elementary Particle
Physics}, ed. N. Svartholm (Almqvist and Wiksells, Stockholm, 1968), p. 367;
S. Glashow, J. Iliopoulos, and L. Maiani,  Phys. Rev. D{\bf 2}, 1285
(1970).

%3
\bibitem{sutheory} 
After several partial and tentative proposals, the
$SU(3)$ strong interaction theory took on its essentially modern form
in: D. Gross and F. Wilczek, Phys Rev. D {\bf 8}, 3633 (1973); Not
coincidentally, the key discovery that allowed one to connect the
abstract gauge theory to experiments, asymptotic freedom, was first
demonstrated just prior to these papers, in: D.~Gross and F. Wilczek,
Phys. Rev. Lett. {\bf 30}, 1343 (1973); H. D. Politzer,
Phys. Rev. Lett. {\bf 30}, 1346 (1973).


%4
\bibitem{ewfig}
LEP Electroweak Working Group, preprint CERN-PPE/96-183 (Dec. 1996).


%5
\bibitem{qcdfig}
M. Schmelling, preprint MPI-H-V39, hep-ex/9701002.  Talk given at the
28th International Conference on High-energy Physics (ICHEP 96),
Warsaw, Poland, 25-31 July 1996.


%6
\bibitem{georgi74}
H. Georgi and S. Glashow, Phys. Rev. Lett. {\bf 32}, 438 (1974).

%7
\bibitem{georgi75}
H. Georgi, in {\it Particles and Fields -- 1974}, ed. C. Carlson
(AIP press, New York, 1975).

%
%\bibitem{unification}
%See S.~Dimopoulos, S.~Raby and F.~Wilczek, Phys. Today {\bf44}, 25 (1991) 
%and references contained therein.

%8
\bibitem{nielsen81}
N. Nielsen, Am. J. Phys. {\bf 49}, 1171 (1981); R. Hughes, Nucl. Phys. B
{\bf 186}, 376 (1981).

%9
\bibitem{quinn74}
H. Georgi, H. Quinn, and S. Weinberg, Phys. Rev. Lett. {\bf 33}, 451 (1974).

%10
\bibitem{blewitt85}
See for example G. Blewitt, {\it et al}, Phys. Rev. Lett. {\bf 55}, 
2114 (1985), and 
the latest Particle Data Group compilations.


%11
\bibitem{ferrara86}
A very useful introduction and collection of basic papers
on supersymmetry
is S. Ferrara, {\it Supersymmetry\/} (2 vols.) (World Scientific,
Singapore 1986).   Another excellent standard reference
is N.-P. Nilles, Phys. Reports {\bf 110}, 1 (1984).
See also [26].

%12
\bibitem{dimopoulos81}
S. Dimopoulos, S. Raby, and F.~Wilczek, Phys.
Rev. D {\bf 24}, 1681 (1981).

%13
\bibitem{ellis91}
J. Ellis, S. Kelley, and D. Nanopoulos,
Phys. Lett. {\bf B260}, 131 (1991);
U. Amaldi, W. de Boer, and H. Furstenau, 
Phys. Lett. {\bf B260}, 447 (1991); for more recent analysis see
P. Langacker and N. Polonsky, Phys. Rev. D {\bf 49}, 1454 (1994).

%14
\bibitem{dienes}
K. Dienes and C. Kolda, IASSNS-HEP-97-68, hep-ph/9712322 (Dec 1997).

%15
\bibitem{swein89}
S. Weinberg,  Rev. Mod. Phys. {\bf 61}, 1 (1989).

%16
\bibitem{topquark}
The existence of the infrared fixed point was first discussed in:
C.~Hill, Phys. Rev. D {\bf 24},  691 (1981). More recent examinations 
including supersymmetry appear in:    V. Barger, M. Berger and P.
Ohmann, Phys. Rev. D {\bf 47}, 1093 (1993);  P. Langacker and N. 
Polonsky, Phys. Rev. D {\bf 47}, 4028 (1993); M. Carena, S.~Pokorski 
and C.~Wagner, Nuc. Phys. B {\bf 406}, 59 (1993).

%17
\bibitem{gell-mann79}
M. Gell-Mann, P. Ramond, and R. Slansky, in 
{\it Supergravity}, ed. P. van Neiuwenhuizen and D. Freedman 
(North Holland, Amsterdam, 1979),
p. 315; T.~Yanagida, Proc. of the Workshop on Unified Theory and
Baryon Number in the Universe, eds. O. Sawada and A. Sugamoto (KEK, 1979).

%18
\bibitem{jjvr}
J. Christenson, J. Cronin, V. Fitch and R. Turlay,
Phys. Rev. Lett. {\bf 13}, 138 (1964).
 
%19
\bibitem{kobay}
M. Kobayashi and T. Maskawa, Prog. Theor. Phys. {\bf
49}, 652 (1973).

%20
\bibitem{nirquinn}
Y. Nir and H. Quinn, Ann. Rev. Nucl. Part. Sci. {\bf 42}, 211 (1992).

%21
\bibitem{barr93}
S. Barr, Int. J. Mod. Phys. {\bf A 8}, 209 (1993).

%22
\bibitem{fw85}
F. Wilczek, The $U(1)$ Problem: Instanton, Axions, and Familons, in
{\it How Far Are We from the Gauge Forces}, ed. A. Zichichi (Plenum,
1985).

%22a
\bibitem{adel91}
E. Adelberger, B. Heckel, C. Stubbs, and W. Rogers,
Ann. Rev. Nucl. Part. Sci. {\bf 41}, 269 (1991). 

%23
\bibitem{cosmic}
The principles are explained in C. Jones, A. Melissinos, {\it Cosmic
Axions}, (World Scientific, 1990).

%24
\bibitem{gjung}
G. Jungman, M. Kamionkowski, K. Griest, Phys. Reports {\bf 267}, 195 (1996).


%25
\bibitem{cohen}
A. Cohen, D. Kaplan and A. Nelson, Ann. Rev. Nucl. Part. Sci. {\bf
43}, 27 (1993).


%26
\bibitem{nature}
For a gentle introduction, see F. Wilczek, Colour Takes The Field,
Nature {\bf 390}, 659 (18/25 Dec. 1998).




\end{thebibliography}
\end{document}

\title{DISCUSSION I}

\author{CHAIRMAN: Professor F. Wilczek}

\maketitle

\begin{center}Scientific Secretaries: A. Adamantini, D. Ghilencea,\\
and R. Harlander
\end{center}

\paragraph{Shovkovy:}
	As you demonstrated, the running couplings in the Standard
Model do not quite match at high energy, but why should we expect a
perfect match?

\paragraph{Wilczek:}
	We do not expect a perfect match.  In fact, if you extrapolate
the minimal supersymmetric extension of the Standard Model (MSSM) with
all the superpartner masses set equal, say to 100 Gev, then the
couplings do not quite meet.  Putting it another way, if you take the
weak and electromagnetic couplings as known, and demand unification
(with the SU(5) normalization of hypercharge) then the strong coupling
-- that is, the $\alpha_s$ evaluated at the $Z$ mass is calculated to
be a bit large, roughly .13 compared to the experimental value .118
$\pm$ .002.  Fortunately, there are known corrections to the simple
extrapolation which are plausibly of the right size.  These come from
the fact that the straight inverse logarithmic running of the
coupling, which is valid for massless particles, is modified for
massive particles.  At a given scale, particles whose masses are much
larger than the scale do not contribute significantly to the vacuum
polarization, and very light particles can be regarded as massless.
However particles with masses comparable to the scale must be handled
more carefully.  The procedure is straightforward in principle,
although it can get messy in detail.

       We expect that there is considerable complexity in the spectrum at two
places, at least: at the electroweak symmetry breaking scale, and at
the unification scale.  The incomplete symmetries of the mass spectrum
at these scales means that the running of the coupling, as it is
affected by these masses, does not obey the symmetries one would have
if the masses were negligible.  One has what are called ``threshold
corrections'' to the naive extrapolation.

	Unfortunately, we do not have a reliable theory of the
spectrum, either at the electroweak scale or the unification scale.
The best we can do, at present, is to try different possibilities --
that is, concretely, different Higgs multiplets, vacuum expectation
values, and soft supersymmetry breaking terms -- to get a rough idea
what the magnitude of the corrections could be, within some class of
models that seem reasonable.  This is a somewhat subjective process --
what seems reasonable to me might not to you -- but the general
consensus is that threshold corrections, especially at the unification
scale, are quite adequate to accommodate the slight discrepancy
between the naive extrapolation of running couplings and experiment.
If we are bold, we can try to turn this argument around, and learn
something about the pattern of masses near the unification scale by
demanding accurate meeting of the couplings at low energy.

	On the other hand, extrapolation of the ordinary Standard
Model, without supersymmetry, differs from experiment by an amount
that cannot be easily accommodated with minor threshold corrections.
One must drastically alter the theory.  Of course, since we are only
fitting one or two parameters, it is possible to find alterations that
will do the job.  But it would be a cruel joke on the part of Nature,
to fool us this way.

\paragraph{Zichichi:} 
       May I make a comment, right away: the story of MSSM is
incomplete, because, after all, it is not enough for them to meet - this
is geometry - afterward they have to continue together. This is why
SU(5) is needed. Please point out that in order to have unification
you need a higher group.

\paragraph{Wilczek:} 
       This is an important additional point that I did not stress
in the lecture. I drew the graphs for running of the couplings in an
oversimplified way, just extrapolating the renormalisation group
evolution linearly in the inverse logarithm, as if the symmetry
continued to be broken past the unification scale. In a more refined
treatment, the symmetry is restored at that scale, and some particles
that were very massive before can no longer be neglected, which
changes the answer.  Clearly symmetry dictates that the couplings will
then all run together.

\paragraph{Giusti:}
	What is the value of the supersymmetric scale you can have
without losing unification of the gauge couplings?

\paragraph{Wilczek:}
	A very important motivation for supersymmetry is that it
protects us from very large radiative corrections to, if you like, the
mass of W or perhaps better, the mass of the Higgs particle. There is
a very dangerous radiative correction in the Standard Model to the
mass term of the scalar Higgs field.  This radiative correction is
quadratically divergent.  Formally, one can add a counter term to
cancel off the divergence.  If one had no expectation for where
essentially new physics beyond the Standard Model occurs, it might
appear reasonable simply to adjust the counterterm so that the
symmetry breaking scale associated with the Higgs field vacuum
expectation value comes out right, following the prescription that the
senseless (infinite) correction must somehow be zero.  But with
unification, there really is another scale that seems quite relevant,
the unification scale, which is something like $10^{15} {\rm Gev}$.  I think it
is quite fair to estimate the radiative correction by putting in the
unification scale for the ultraviolet cut-off; there is no known way
to protect structure at that scale leaking down in this way.  Of
course, if we do this, we get a ridiculously large correction to the
mass term, much larger than what we want the total answer to be.
This, in a nutshell, is the naturality problem associated with the
quadratic divergence of radiative corrections to the Higgs mass.

	It is a very great virtue of supersymmetry that for every
bosonic virtual exchange contributing to radiative corrections of this
kind, there is a fermionic exchange contributing with the opposite
sign which would, if the masses of the boson and fermion were equal,
exactly cancel it. If the masses of these particles are not exactly
equal, their contributions do not cancel exactly, but only
approximately. It will (because of the underlying quadratic
divergence) go like the square of the mass difference, at least for
large splittings. Now, if we want the squared mass parameter of the
Higgs particle to be of the order of $250~ {\rm GeV}^{2}$, and the radiative
contribution to be not much larger than the total answer, then we have
to say that the square of the mass of the supersymmetric partners of
the known Standard Model particles times the fine structure constant
(perhaps divided by $\pi$) is not much greater than this.  In this
simple-minded way, we find strong motivation to assume that the mass
of the supersymmetric particle is of the order of a few TeV. It could
be less, of course, but at most it should be a few TeV.  So, now
coming back to the unification calculation, we can ask if we vary the
masses from nothing to a few TeV, does the calculation stay adequately
accurate. The answer is that it works quite well, as I have already
mentioned.  Perhaps superpartner masses on the high side are favored,
but it is really within the uncertainties due to possible threshold
effects on the other unification side.  On the other hand if we drive
the masses orders of magnitude larger than this it would start to
upset the agreement of the running of the couplings with
experiment. So, you have to take the whole package, with low-energy
supersymmetry -- asymptotic (e.g., Planck scale) supersymmetry is not
good enough.  Both in protecting the weak scale and in making the
couplings unify correctly, supersymmetry can only do its job if the
scale for finding visible supersymmetric partners is a few Tev or
less.  This should be very encouraging for anyone involved in
experimental work at LHC, for example.

\paragraph{Buchel:}
	Is it fair to say that when people are writing unification of
the gauge couplings, they usually do one loop running?

\paragraph{Wilczek:}
	The calculations have been done certainly to two-loop and I
think maybe even to three-loop order.

\paragraph{Buchel:}
	My second question: can you discuss the status of the Yukawa
coupling unification? One would like to push these ideas further.

\paragraph{Wilczek:}
	I think it is appropriate to emphasize gauge coupling
unification because gauge couplings are related to the central parts
of the theory, the parts that we understand profoundly.  Also, the
relevant predictions concern are gauge couplings which have been
measured very well, so there can be are reliable and robust tests of
the basic ideas.

	Emboldened by success for the gauge couplings, one can become
more ambitious and ask about Yukawa couplings and their running.  Any
such discussion has to begin with a big caveat: if the Higgs sector is
complicated, containing several fields, then we cannot from what is
known at present determine all the Yukawa couplings and their
expectation values separately. A unique relationship between what we
know -- namely the particle masses and mixing angles -- and the size
of the Yukawa couplings, arises only if there is a single Higgs
field. For then the masses are equal to a single universal vacuum
expectation value multiplied by the corresponding Yukawa couplings.
The vacuum expectation value is fixed by the observed mass of the W
boson, because in that case we know both the mass and the (gauge)
coupling.  In general, however, we cannot pin things down, because the
W mass fixes only one combination out of many possible vacuum
expectation values. For instance in the MSSM one needs two Higgs
fields, and the ratio of their vacuum expectation values is an unknown
parameter in supersymmetric model building, labelled $\tan \beta$.

	Keeping that in mind, we can still ask: is there anything in
the observed spectrum that suggests the relevance of the running of
the Yukawa coupling? I would say there are two hints, though I must
emphasize again that they are by no means on the same footing as
unification of gauge couplings. One suggestion is that the ratio of
quark to lepton masses for particles in the same family evolves by
renormalizing from a value of unity at the unification scale to what
we observe.  This equality at the unification scale can arise for
group-theoretic reasons in a unified theory such as SU(5), where
quarks and leptons are put in common multiplets.  If you try this idea
you obtain a pretty good value for the mass of the b-quark divided by
the mass of the tau.  Taking this idea to its logical conclusion, one
would predict a universal ratio between the mass of the
charge-(-1/3)-quark and the mass of the corresponding negatively
charged lepton in each family, which does not work in the other cases.
So something is missing.  There has been a lot of work to suggest what
it is, but no consensus has yet emerged.

	The other, and I think much more profound, suggestion for
relevance of running ideas to masses has to do with the top quark
mass.  Although we certainly do not understand the masses of all the
quarks and leptons, there is one whose mass might be much simpler to
understand than all the others, and that is the top quark. People
sometimes say that the top quark is anomalously heavy but actually,
from a higher perspective, it is the only known elementary fermion
that has a reasonable mass. $M_{t}$ is the Yukawa coupling times the
expectation value of the Higgs field. In the minimal model Standard
Model this expectation value of the Higgs field is 250 GeV.  This is
determined from the W mass.  Given this, one sees that it is only for
the top quark that the Yukawa coupling is at all close to one. In all
other cases it is very very small. For the electron, e.g., it is about
a part in a million. This raises the possibility that the top quark
mass is governed by an infrared fixed point. That is, it might be that
the top quark mass is understood in the following way: no matter what
the Higgs coupling is at the unification scale, when we run it down we
find there is a focusing effect, implying that more or less any large
value of the Yukawa coupling at the unification scale would run down
to something close to the observed one at low energies. This running,
of course, is only important for Higgs couplings that are strong at
unification; very small values stay small.  In models with more than
one Higgs field its is conceivable that this mechanism applies to
other fermions as well.  The most plausible possibility, perhaps, is
that the b quark sits at an infrared fixed point; this can arise in
supersymmetric models with large $\tan \beta$.

\paragraph{Armoni:}
	At the beginning of your lecture you mentioned the basic
principles of the SM, like unitarity and locality. It is believed that
in quantum gravity at least one of these principles must be
abandoned. Is there any proof of the possibility of having some
extended gravity which is given in terms of local quantum field
theory? Is it theoretically excluded that we can have a local quantum
field theory including gravity?

\paragraph{Wilczek:}
	No, I think it is not excluded at all. In my opinion the jury
is still out on quantum gravity.  But one cannot deny that the
conceptual world of gravity is quite different from the usual
conceptual world of local quantum field theory. For example, already
using the word locality assumes that you have some notion of nearness,
and if your basic symmetry is just diffeomorphism invariance, as it is
perhaps in quantum gravity -- if the metric is a derived concept --
then that just does not make sense. But maybe the metric is not a
derived concept, we just do not know. To say something constructive,
let me address the situation in string theory, which is the closest
thing we have at present to a quantum theory of gravity. Despite some
prominent claims in the literature, it is very far from obvious to me
that string theory is non-local in any real sense. In string theory,
you really have interactions just when the strings touch, and the
effects of this interaction travel along the strings at a speed
limited by the speed of light, so I believe that at least
perturbatively the theory is local.

	It was one of the great achievements of Galileo to realize
that science is local, that you do not have to worry very much for
experiments here about what is going on on the far side of the moon,
that astrology is misguided, and so forth. These are assumptions that
are implicit in almost all scientific work, and if there were really
significant non-locality one might have expected something to go
terribly wrong in all this work by now.  So I am a bit nervous about
giving up locality; but I am also not sure that it is given up in
string theory.  I am not aware anybody who has pointed at a specific
non-local effect and made a testable prediction, even for a thought
experiment or impractical observable.  Without that, the discussion is
metaphysical.

\paragraph{Ryan:}
	I have to complain a bit about the presentation of the running
of the strong coupling. A naive student might come away with the
impression that we calculate the coupling constant completely from
first principles and that the data points fall dutifully on top of the
curve. Since this is a School, I think it should be emphasized that
the evolution equations describe the shape of the curve, but the
absolute magnitude must be derived from fits to the data points.

\paragraph{Wilczek:}
	We can look at the data, as presented in the lecture, and
address this point. The point I am making about the absolute coupling
is simply that the curve giving the effective strong coupling is
highly focusing.  More quantitatively, the phenomenon is as follows.
In solving the renormalization group equation for the running of the
coupling, a mass parameter called lambda appears.  Roughly speaking,
it sets the scale at which the effective coupling becomes large.
Lambda has to be derived from measurements; as you say, therefore, the
prediction is a one-parameter set of curves. However, if lambda is
chosen to be any typical strong interaction scale, ranging from, say,
100 MeV to 1 GeV, any specific choice will give you almost identically
the same value of the effective strong coupling at scales of a hundred
GeV. In that sense, this coupling is given with about 10 per cent
accuracy as a no-parameter prediction of the theory.

\paragraph{Zichichi:}
	Do you place lattice QCD at the same level as an experimental finding?

\paragraph{Wilczek:}
	Lattice QCD is not an abstract game; it is a well-defined
method for calculating consequences of a well-defined theory.  So it
is as legitimate to compare its results with experiments, as to
compare the results of perturbative QCD with experiment.  In practice,
the determination of the strong coupling from using lattice QCD is
based on fitting its results to the experimental measurement of heavy
quark spectra. It is a comparison of theory and experiment, as are all
the other points on the curve, so it is really no different in
principle.  By the way, many of these points summarize tens or
hundreds of independent measurements under different
circumstances. Any of them could have falsified QCD, but none has.

\paragraph{Paus:}
	I might make a comment on your plot. There are quite a few
measurements missing, e.g., LEP now measures the strong coupling at
183 GeV and still theory and experiment are in very good agreement.

\paragraph{Wilczek:}
	Yes, thank you. I do not think the essential correctness of
QCD, and of our methods of calculation, are negotiable items anymore.

	Let me add one more remark: when we defined `years'
of particle physics, one of the remarkable things taken to mark the
beginning was the Lamb shift; that supplied convincing evidence for
virtual particles and the correctness of quantum field theory. The
features of heavy quark spectra that are calculated by combining
perturbative calculations and lattice QCD, and used in determining the
strong coupling, are the exact analogs to the Lamb shift. In fact, one
of them is just the strong interaction version of Lamb shift.

\paragraph{Roos:}
	Your classification of states in a 16-dimensional
representation of SO(10) made use of a notation with five signs,
pluses and minuses. It struck me that this representation did not
exhaust all the permutations of signs. Those which are absent, do they
live in another representation or are they forbidden by some
principle?

\paragraph{Wilczek:}
	There is a rule here that you have to have an even number of
pluses. That is to make the representation irreducible. By the way,
the ones with the same number of pluses are the SU(5)
multiplets. There are ten entries with two plus signs which fill out
the antisymmetric tensor of SU(5), five entries with four plus signs,
those being the 5 of SU(5), and then there is the singlet.

\paragraph{Bruyninckx:}
	What is your opinion about what the QCD phase diagram should look like?

\paragraph{Wilczek:}
	I think I have understood for a long time what it looks like,
as a function of the temperature, and even maybe temperature and
strange quark mass, at zero chemical potential.  This is a little
complicated to discuss here, but well documented in the literature.
On the other hand adding a chemical potential, as is necessary to
discuss high-density situations like neutron stars or certain aspects
of heavy ion collisions, rapidly brings us into poorly explored
territory.  I will say a little about it tomorrow. It will be some
time before the dust settles on these questions, in my opinion.

\title{DISCUSSION II}

\author{CHAIRMAN: Professor F. Wilczek}

\maketitle

\begin{center}Scientific Secretaries: K. Riesselmann, S. Tapprogge,\\
 and A. Wodecki

\end{center}

\paragraph{Ghilencia:}
	We had a long discussion on the so-called threshold effects
and the two-loop corrections in the running of the gauge couplings. I
just want to add that these threshold effects, considered in their one
loop approximation - which is consistent for a two loop running - are
cancelled by the two loop terms in the beta function. This greatly
simplifies the renormalization group equations, which, interestingly
enough, get a one-loop-like form if expressed in terms of wave
function renormalization coefficients.

\paragraph{Ghilencia:}
	Do you see any physical meaning of the Grassman variables we
integrate over, as they are, to my knowledge, introduced only on
mathematical grounds?

\paragraph{Wilczek:}
	Grassmann variables are an implementation of Fermi statistics.
I do not know of any other physical meaning.  They play an important
role in supersymmetry, of course.

\paragraph{Brown:}
	Are there any more profound, a priori, reasons for an SU(5) or
SO(10) gauge principle?

\paragraph{Wilczek:}
	Well, one widely held speculation is that one should be
working with weakly coupled heterotic strings that are compactified to
four space-time dimensions.  In the overlying ten dimensional theory,
the gauge group is E8 x E8 emerges.  This group naturally contains E6
and hence SO(10), SU(5), and of course the Standard Model gauge group.
A well-studied class of compactifications involves putting a
non-trivial gauge field on the compactified manifold, which reduces
the symmetry of one E8 factor down to E6 or less. Whether something
along these lines describes reality, it is too soon to say.

	Let me make a couple of more naive remarks.  A particularly
intriguing feature of SO(10) is its spinor representation, used to
house the quarks and leptons, in which the states have a simple
representations in terms of basis states labelled by a set of ``+'' and
``-'' signs.  Perhaps this suggests composite structure. Alternatively,
one could wonder whether the occurrence of spinors both in internal
space and in space-time is more than a coincidence.  These are just
intriguing facts; they are not presently incorporated in any
compelling theoretical framework as far as I know.

	Here is another interesting question: do spinor
representations occur in quantum mechanical systems other than the
ones connected to the SO(3) rotational group?  Note that the spinor
representation is a projective representation: there exists a phase
which cannot be absorbed into the unitary transformations of the
group.  Non-trivial projective representation also arise for the group
$S_N$ relevant to the quantum kinematics of identical particles, for $N
>3$.  Remarkably, they have the same structure as spinor
representations, including exponential growth with N of the dimension
of the irreducible representation.  So perhaps there is a connection
between exotic quantum statistics and the existence of internal spinor
representations.

\paragraph{'t Hooft:} 
	In your theory of colour-superconductivity with non-zero
expectation value for a diquark state, how is the complete
flavour/colour group $SU(N_{f})_{L} \times SU(N_{f})_{R} \times U(1)
	\times SU(3)_{C}$ broken? 

\paragraph{Wilczek:}
	I have looked most carefully at the case of two light
flavours.  In this case the diquark forms an isosinglet (in fact,
chiral isosinglet) scalar.  With three light flavors, there is a very
interesting and probably energetically favorable possibility to break
the SU(3) x SU(3) of color times flavor to the diagonal subgroup; this
of course breaks both color and strangeness spontaneously.  I have not
thought much about more flavors.

\paragraph{'t Hooft:}
	Will nucleons become massless in this approach?

\paragraph{Wilczek:}
	In the phase where chiral symmetry is restored it is not
useful anymore to think about nucleons because the elementary
excitations are quark excitations which feature a Fermi surface,
generally with a gap.  Its spectrum very much resembles that of a
Higgs phase.  In the 2-flavor case an SU(2) subgroup of the original
SU(3) colour group survives, and it can possibly lead to confinement
of two colors.  There are also, however, colour-neutral quarks.  With
three flavors, and the symmetry breaking pattern I mentioned, there is
no local symmetry left, and presumably no question of confinement.
There is, however, a gap for all elementary excitations, even though
the underlying theory is full of massless particles.

\paragraph{Masiero:}
	This morning you mentioned that you have some reasons for
skepticism towards inflationary scenarios. Can you detail them?

\paragraph{Wilczek:}
	Skepticism is too strong a term; I just think the evidence is
very thin, and in such a situation one can imagine alternatives will
arise.  In any case it is a healthy exercise to take a critical look.
	
   Inflationary models clearly leave room for improvement with respect
to fundamental physics even though there are - to my knowledge - no
serious cosmological objections to inflationary models.  First, there
is the cosmological term problem.  From experiment we know both that
$\Lambda=0$, very nearly, and that the structure of the present-day vacuum
state is quite complicated.  Inflation, however, depends on $\Lambda$ sitting
far from zero for a long period of time when the vacuum was
simpler. This is scary.  Second, it is disturbing that the postulated
inflation field is not connected to a fundamental theory or microscopic
model. And third, predictions from inflation are mostly of a rather
generic character, and not very characteristic.  The predictions of a
flat universe and a nearly flat fluctuation spectrum, for example,
were already proposed by Harrison, Zeldovich and others based on
naturality and stability arguments before inflation was proposed.
Hence inflation seems to provide largely ``postdictions'' of qualitative
features that - because they are simple - might have alternative
explanations.

\paragraph{Veneziano:}
	My strong support for SUSY is wearing thin. On one side, SUSY
helps to understand hierarchy, the mass of the Higgs boson and grand
unification. However the standard model - which SUSY should replace -
contains automatically the conservation of lepton and baryon quantum
numbers, which has to be introduced in a complicated way in SUSY
models, so as not to be in contradiction with existing data. What is
your honest opinion about SUSY?

\paragraph{Wilczek:}
	I find the unification of couplings, deduced in a
straightforward way from an unforced fit, most convincing. If this
were to prove a mere coincidence, it would be a cruel joke on the part
of the Creator.  There are several other nice features of
supersymmetry, including its ability to protect the weak scale as I
discussed yesterday, its general helpfulness in reducing ultraviolet
divergences, its potential to unify gauge fields and matter fields --
both of which we know to exist, and its connection to string theory.
But these are rather generic, like the evidence for inflation.  To me,
the unification of couplings is on a different plane, and provides by
far the best reason to believe in SUSY.

\paragraph{Veneziano:}
 	SUSY might be the only chance to understand why the
cosmological constant is 0 (by leaving it 0 during the breaking of
SUSY).

\paragraph{Wilczek:}
	Maybe, but in the present state of understanding this is
problematic, on at least two grounds.  First, in supergravity, as
opposed to global supersymmetry, the connection of energy to
cosmological term is not so direct, and it is not necessarily so that
the ground state of unbroken supersymmetry occurs with zero
cosmological term -- indeed, in some ways supersymmetry is happiest in
anti-deSitter space.  Second, we know that in reality supersymmetry is
broken, and that the vacuum energy density associated with the
breaking, at least if we focus attention on the manifest low-energy
degrees of freedom, exceeds by many orders of magnitude the upper
limit on the cosmological term.  So if supersymmetry is to address the
cosmological term problem, it will need help from additional ideas.